\documentclass[sigplan,screen]{acmart}

\usepackage[utf8]{inputenc} 
\usepackage{url}

\usepackage[linesnumbered,ruled,vlined]{algorithm2e}
\usepackage{algpseudocode}

\usepackage{cleveref}

\usepackage{cancel}
\usepackage{changepage}
\usepackage{caption}
\usepackage{threeparttable}
\usepackage[inline]{enumitem} 
\usepackage{color, soul} 

\usepackage{amsmath,amssymb,amsfonts}
\usepackage{graphicx}
\usepackage{textcomp}
\usepackage{xcolor}

\usepackage{booktabs} 

\newcommand{\ciclad}{$\textit{Ciclad}$}
\newcommand{\cicladA}{$\textit{Ciclad}^{+}$}
\newcommand{\cicladD}{$\textit{Ciclad}^{-}$}

\newcommand{\moment}{$\textit{Moment}$}
\newcommand{\newmoment}{$\textit{NewMoment}$}
\newcommand{\cfistream}{$\textit{CFI-Stream}$}
\newcommand{\galiciap}{$\textit{Galicia-P}$}
\newcommand{\galiciam}{$\textit{Galicia-M}$}
\newcommand{\galiciat}{$\textit{Galicia-T}$}
\newcommand{\clostream}{$\textit{CloStream}$}

\newcommand{\gctree}{$\textit{GC-Tree}$}
\newcommand{\dciclosed}{$\textit{DCI-Closed}$}

\def\newdef#1#2{\newtheorem{definition}{Definition}}

\newdef{definition}{Definition}
\newtheorem{theorem}{Property}
    
\renewcommand\footnotetextcopyrightpermission[1]{}
\begin{document}
\fancyhead{}
\title{CICLAD: A Fast and Memory-efficient Closed Itemset Miner for Streams}
\author{Tomas Martin, Guy Francoeur, Petko Valtchev}
\affiliation{%
\institution{Centre de recherche en intelligence artificielle (CRIA), UQÀM}
\city{Montréal}
\state{Québec, Canada}
}
\date{}

\begin{abstract}
Mining association rules  from data streams is a challenging task due to the (typically) limited resources available vs. the large size of the result. Frequent closed itemsets (FCI) enable an efficient first step, yet current FCI stream miners are not optimal on resource consumption, e.g. they store a large number of extra itemsets at an additional cost.  
In a search for a better storage-efficiency trade-off, we designed \ciclad, an intersection-based sliding-window FCI miner. Leveraging in-depth insights into FCI evolution, it combines minimal storage with quick access. Experimental results indicate \ciclad's memory imprint is much lower and its performances globally better than competitor methods.
\end{abstract}

\maketitle


\section{Introduction}

Association rule (AR) and the component frequent itemset (FI) mining from data streams have found a wide range of practical applications, e.g. in network traffic analysis, click stream mining, online transaction analysis~\cite{rashid2013mining,hamadi_2016_compiling, karim2018mining}. 

Both task are challenging due to the specific conditions of the stream environment: dynamic data inflow, potential concept drift and, typically, limited resources~\cite{yamamoto2019parasol, jiang2006research}. Therefore, the design of efficient stream miners has to reflect concerns such as single access to stream data, compact storage of results, resource limit-awareness, etc. 
Moreover, various processing models have been investigated to reflect decay of data in the stream, such as \textit{landmark}, \textit{sliding} or \textit{damped} window.
In a different vein, while a large number of stream miners target interesting itemsets (\textit{frequent} ~\cite{li2009mining}, \textit{rare}~\cite{huang2012rare}, \textit{closed}~\cite{chi2004moment}, \textit{maximal}~\cite{karim2018mining}, \textit{self-sufficient}~\cite{tang2019adaptive}, etc.), only few methods would maintain the associations, arguably due to the huge number of rules to process on each window shift.

Our ultimate goal is the design of an efficient AR stream miner, whereby we propose to tackle the inherent complexity of the task by reducing the output to some concise representation of strong rules, e.g. as described in~\cite{marzena02}. Such representations typically constructed on top of distinguished itemsets such as closed itemsets (CIs), generators (a.k.a. \textit{free} itemsets), or maximal FIs~\cite{cond-rep:calders+04}, and might exploit additional structure on those, e.g. lattice links~\cite{zaki2005efficient}.
A variety of batch methods exist for these representations (e.g. see~\cite{marzena02,cond-rep:calders+04}), yet to the best of our knowledge they have not been studied in stream settings.
Stream miners have been designed for (frequent) CIs~\cite{chi2004moment,li2006new,valtchev02,jiang2006cfi,Valtchev2008,yen2011fast} as well as for frequent generators~\cite{gao2009efficient}.
The former split into two categories: Methods~\cite{chi2004moment,li2006new} adapt batch pattern enumeration~\cite{zaki97}, as opposed to those relying on CI intersection~\cite{valtchev02,jiang2006cfi,Valtchev2008,yen2011fast}.
Intersection-based closure computing is rooted in formal concept analysis (FCA)~\cite{fca_book}, where a variety of incremental, i.e. landmark, miners target closed itemsets (CIs) (dating back to~\cite{galois_inc:godin+95}). Moreover, various derived problems of interest, e.g. mining  CIs with links and generators, have been studied~\cite{pfaltz02b}. However, resource-awareness is not a prime concern in FCA, and neither is data decay, hence no decremental methods have been designed (to achieve a sliding window mode).

In this paper, we focus on the groundwork task to support the design of fully-blown concise AR stream miner, i.e. the online mining of FCIs. Our analysis of the literature indicates that all existing methods have increased memory consumption as they need to store extra itemsets on top of the target FCIs: The first group requires few infrequent and large number of non closed itemsets whereas the second one maintains all infrequent CIs. However, pattern enumeration miners additionally maintain tidsets, which is costly, especially on large or highly dynamic streams, whereas intersection-based ones skip tids altogether.

We chose to follow an intersection-based approach which, despite the aforementioned overhead due to infriquent CI maintenance, offers distinct advantages such as higher flexibility (e.g. upon support threshold decreases) and versatility (both frequent and rare patterns targeted).
Moreover, the CI indexing approach used in both~\cite{Valtchev2008,yen2011fast} seems particularly appealing, yet both methods suffer on superfluous storage and/or processing.

As a remedy, we propose \ciclad, a two-fold CI stream miner whose incremental part \cicladA~streamlines quick access to CIs and item-wise intersection growth from~\cite{Valtchev2008} while skipping non essential storage of itemsets. The decremental \cicladD, in turn, implements some novel insights into CI evolution that allow it to fit the same overall computing schema as \cicladA. The resulting homogeneous compound method achieves both high efficiency and low memory usage. This has been confirmed by a validation study over both real and synthetic datasets (retail, network security, etc.) whose outcome shows that \ciclad~outperforms competing methods by a comfortable margin on all but the most dense datasets.

Our paper's contributions are as follows: 
(1) formalization of transaction removal, 
(2) 
a unified intersection-based sliding-window miner \ciclad,
(3) performance study on a variety of CI mining methods for streams (our code and datasets publicly available\footnote{\url{https://github.com/guyfrancoeur/ciclad}}). Additionally, we provide correctness proofs enhancing~\cite{Valtchev2008} (see Appendix).

In what follows, section~\ref{sec:bgd} provides background on CIs and online mining while section~\ref{sec:related} summarizes previous work. Our mathematical approach and the algorithmic details of \ciclad~are presented in sections~\ref{sec:design} and~\ref{sec:design-code}, respectively. Section~\ref{sec:eval} summarizes the performance study. Concluding remarks are given in section~\ref{sec:conclude}.

\section{Background}
\label{sec:bgd}

Below, we recall basics of pattern mining, closed patterns and stream mining.

\subsection{Frequent (closed) itemsets}
\label{sec:bgd-fci}

Assume a \emph{transaction database} ${\mathcal D}$ (as in Table~\ref{tab:example_data}) defined on top of a set of \emph{items} ${\mathcal I} =\{a_1,a_2,\dots, a_n\}$ (here $\{a, \dots, h \}$). A set $X \subseteq {\mathcal I}$ is called an \emph{itemset} while a transaction is a pair ($tid$, itemset) where $tid$ is a \emph{transaction identifier}. Similarly, a set of tids is called a \emph{tidset}.

\begin{center}
\begin{tabular}{|l||l||l|}
\hline
    
   	($\bar{1}$, $abcdefgh$) & ($\bar{5}$, $g$) & ($\bar{9}$, $d$)  \\
   	($\bar{2}$, $abcef$) & ($\bar{6}$, $efh$) & ($\bar{10}$, $bcgh$)\\
   	($\bar{3}$, $cdfgh$) & ($\bar{7}$, $abcd$) &  \\
   	($\bar{4}$, $efgh$) & ($\bar{8}$, $bcd$) &  \\
\hline
\end{tabular}
\captionof{table}{Sample data, further referred to as \textbf{${\mathcal D}_{10}$}}
\label{tab:example_data}
\end{center}

Given a tidset $Y$ from $\mathcal{D}$, $\iota_{\mathcal{D}}(Y) = \bigcap \{Z | (j,Z) \in \mathcal{D}, j \in Y\}$ denotes the itemsets shared by the respective transactions. Conversely, the \emph{support set} of an itemset $X$ comprises tids whose itemsets cover $X$, $\tau_{\mathcal{D}}(X)= \{j | (j,Z) \in \mathcal{D}, X \subseteq Z\}$. For instance,  $\tau_{\mathcal{D}}(ab)=\{\bar{1},\bar{2},\bar{7}\}$ and $\iota_{\mathcal{D}}(\{\bar{2},\bar{7} \})= abc$. For a tid $j$ from $\mathcal{D}$, $\iota(j)$ denotes its itemset: $\iota(j) = Z$ iff $(j,Z) \in \mathcal{D}$. To shorten notations, we omit the subscript $_{\mathcal{D}}$ (if no confusion is possible) and use $t_*$ to denote $t_*$'s itemset.

Quality of an itemset $X$ follows the size of its support set, a.k.a. its \textit{support} $\sigma(X)= |\tau(X)|$. 

A binary \textit{frequency} criterion is based on a pre-defined \emph{minimum support} threshold, or \emph{min\_supp}, denoted $\varsigma$. The ensuing family of frequent itemsets in $\mathcal{D}$ will be $\mathcal{F}(\mathcal{D}, \sigma)$.
Support sets induce an equivalence relation on $\wp({\mathcal I})$: $X \cong Z$ iff $\tau(X)=\tau(Z)$ (e.g. $ab \cong ac$).
The equivalence class of $X$, $[X]_{\mathcal{D}}$ admits a
unique maximum, a.k.a. \emph{closed} itemset (CI) which,
following the anti-monotony of $\sigma$, can be defined as follows:
\begin{definition}
\label{def:closed}
$X \subseteq {\mathcal I}$ is
\emph{closed}
if no proper superset thereof has the same support.
\end{definition}
In $\mathcal{D}$, $abc$ is closed while $b$ is not ($\tau(bc)=\tau(b)$).
The CIs of $\mathcal{D}$ will be denoted $\mathcal{C}(\mathcal{D})$ and frequent ones $\mathcal{FC}(\mathcal{D}, \varsigma) $.

In formal concept analysis (FCA)~\cite{fca_book}, CIs, termed \textit{concept intents}, are defined via a \emph{closure} operator $\kappa$ induced by $\mathcal{C}(\mathcal{D})$, $\kappa : X \mapsto \max([X]_{\mathcal{D}})$, hence $X=\kappa(X)$ iff $X$ is closed. Here, $\kappa$ is nothing more than the composition of $\iota$ and $\tau$:
\begin{theorem}
\label{prop:fca-fci}
Given a $X \subseteq {\mathcal I}$, $\kappa(X)= \iota(\tau(X))$.
\end{theorem}
%
Here, $\kappa(ab) = abc$ (as $\iota(\tau(ab)) = \iota(\{\bar{1},\bar{2},\bar{7}\}) = abc$ and $abc = \max([ab]_{\mathcal{D}})$).
Next, CIs are exactly the intersections of arbitrary sets of transactions~\cite{fca_book}.
\begin{theorem}
\label{prop:fca-transactions-int}
$\mathcal{C}(\mathcal{D}) = {\mathcal D}^{\cap}$.
\end{theorem}
%
For example, in
$\mathcal{C}(\mathcal{D})$ as given in Table~\ref{tab:ci_before_rem_second}, 
$ef$ (CI $6$), can be generated as $6 = \bar{2} \cap \bar{6}$, whereas  $\bigcap\{2, 15, 22\}$ yields $bc$ (CI $16$). As a corollary, $\mathcal{C}(\mathcal{D})$ is closed under $\cap$ (and so is $\mathcal{FC}(\mathcal{D}, \varsigma)$). 

\begin{center}
\begin{tabular}{|c|l||c|l||c|l|}
\hline
   	$1$ & $(abcdefgh:1)$ & $9$ & $(g:5)$     & $16$ & $(bc:5)$\\
   	$2$ & $(abcef:2)$    & $10$ & $(fh:4)$   & $17$ & $(bcd:3)$\\
   	$3$ & $(cf:3)$       & $11$ & $(efh:3)$  & $18$ & $(d:5)$\\
   	$4$ & $(cdfgh:2)$    & $12$ & $(c:6)$    & $19$ & $(h:5)$\\
   	$5$ & $(f:5)$        & $13$ & $(cd:4)$   & $20$ & $(gh:4)$\\
   	$6$ & $(ef:4)$       & $14$ & $(abc:3)$  & $21$ & $(cgh:3)$\\
   	$7$ & $(fgh:3)$      & $15$ & $(abcd:2)$ & $22$ & $(bcgh:2)$\\ 
   	$8$ & $(efgh:2)$     & & & & \\
\hline
\end{tabular}
\captionof{table}{The family $\mathcal{C}(\mathcal{D}_{10})$ ($\sigma$ values behind ':')}
\label{tab:ci_before_rem_second}
\end{center}

\subsection{Mining FCIs over a stream}
\label{sec:bgd-stream}

Stream pattern mining amounts to updating the pattern family of a window upon adding or removing a transaction (called \textit{increment} and \textit{decrement}, respectively). For instance, assume the transactions in Table~\ref{tab:example_data} are acquired in the order of their tids. Let $t_n$ be a \textit{new} transaction not in $\mathcal{D}$ and let $\mathcal{D}^+ = \mathcal{D} \cup \{t_n\}$.
Simply put, the incremental update results in new CIs being added in $\mathcal{C}(\mathcal{D}^+)$ and $\sigma_{\mathcal{D}^+}$ values computed for them as well as for some existing CIs (support sets extended by $t_n$). Indeed, following Property~\ref{prop:fca-transactions-int}, no CI from $\mathcal{D}$ can vanish in $\mathcal{D}^+$ as $\mathcal{C}(\mathcal{D}^+) = (\mathcal{C}(\mathcal{D}) \cup \{t_n\})^{\cap}$ entails $\mathcal{C}(\mathcal{D}) \subseteq \mathcal{C}(\mathcal{D}^+)$.
As an illustration, assume $\mathcal{D}_{\bar{1},\bar{9}}= \{\bar{1}, \ldots, \bar{9}\}$ (see Table~\ref{tab:example_data}) with its CI family $\mathcal{C}(\mathcal{D}_{\bar{1},\bar{9}})$ as given in 
Table~\ref{tab:ci_before_tenth} and $t_n=\bar{10}$. 
Out of the latter two, the incremental method would output $\mathcal{C}(\mathcal{D}_{10})$ (Table~\ref{tab:ci_before_rem_second}).

\begin{center}
\begin{tabular}{|c|l||c|l||c|l|}
\hline
   	$1$ & $(abcdefgh:1)$ & $7$ & $(fgh:3)$  & $13$ & $(cd:4)$\\
   	$2$ & $(abcef:2)$    & $8$ & $(efgh:2)$ & $14$ & $(abc:3)$\\
   	$3$ & $(cf:3)$       & $9$  & $(g:4)$   & $15$ & $(abcd:2)$\\
   	$4$ & $(cdfgh:2)$    & $10$ & $(fh:4)$  & $16$ & $(bc:4)$\\
   	$5$ & $(f:5)$        & $11$ & $(efh:3)$ & $17$ & $(bcd:3)$\\
   	$6$ & $(ef:4)$       & $12$ & $(c:5)$   & $18$ & $(d:5)$\\
\hline
\end{tabular}
\captionof{table}{The CI family $\mathcal{C}(\mathcal{D}_{\bar{1},\bar{9}})$}
\label{tab:ci_before_tenth}
\end{center}

Algorithm-wise, as shown in~\cite{Valtchev2008}, all \textit{new} CIs from $\mathcal{C}(\mathcal{D}^+) - \mathcal{C}(\mathcal{D})$ are generated as intersections of $t_n$ with a CI from $\mathcal{C}(\mathcal{D})$, e.g. CI $20$ ($gh$) arises as $\bar{10} \cap 8$.
Let $\Delta(\mathcal{D}, t_n) = \{t_n \cap c ~|~ c \in \mathcal{C}(\mathcal{D})\}$
denote the intersection set generated by $t_n$. Here,  
$\Delta(\mathcal{D}_{\bar{1},\bar{9}}, \bar{10}) = \{bcgh, cgh, bc, gh, c, g, h\}$.
Observe that some itemsets correspond to CIs from $\mathcal{C}(\mathcal{D}_{\bar{1},\bar{9}})$, e.g. $bc$ is the CI $16$ in $\mathcal{D}_{\bar{1},\bar{9}}$. We shall call these CIs \textit{promoted}, as opposed to \textit{new} ones, and denote them $\mathcal{C}_P(\mathcal{D})$. It is readily shown that promoted CIs are exactly those included in (the itemset of) $t_n$. Correspondingly, new intersections in $\Delta(\mathcal{D}, t_n)$, denoted $\mathcal{C}_N(\mathcal{D})$, can only involve CIs that are not included in $t_n$. Observe that whether a new or a promoted CI, there may be multiple ways to generate an itemset $X \in \Delta(\mathcal{D}, t_n)$, e.g. $bc$ is also $\bar{10} \cap 2$. In fact, $t_n$ induces an equivalence relation over $\mathcal{C}(\mathcal{D})$ in which a class is defined as $[c]_{t_n} = \{\bar{c}~|~\bar{c} \cap t_n = c~\cap~t_n \}$ for any CI $c$ (e.g. $8 \in [7]_{\bar{10}}$). Clearly, each class is associated with some $X$ from $\Delta(\mathcal{D}, t_n)$, safe the one gathering CIs that are disjoint with $t_n$ (of no interest here). In Figure~\ref{fig:highlevel} (section~\ref{sec:over-ex-inc}), the grey-filled table presents the equivalence classes associated to $\Delta(\mathcal{D}_{\bar{1},\bar{9}}, \bar{10})$.

Crucially, each class $[~]_{t_n}$ associated to some intersection $X$ has a distinguished member CI that \textit{canonically generates} $X$ (CI in bold in Figure~\ref{fig:highlevel}). 
This is the CI corresponding to $\kappa_{\mathcal{D}}(X)$, the closure of $X$ in $\mathcal{D}$, which is, provably, the minimum of the class (see~\cite{Valtchev2008}). 
Now, if $X$ is a promoted CI, then it is closed in $\mathcal{D}$ ($X=\kappa_{\mathcal{D}}(X)$), hence it equals the canonical member ($X = \min_{\subseteq}([X]_{t_n})$. 
Otherwise, $X$ is new CI w.r.t. to $\mathcal{D}$, hence non closed and strictly smaller than its closure ($X \subseteq \kappa_{\mathcal{D}}(X)$) that is further called the \textit{genitor}\footnote{In~\cite{galois_inc:godin+95}, \textit{genitors} are called \textit{generators} and \textit{promoted} - \textit{modified}.} of $X$.
In our example, $gh$ is a new CI in $\mathcal{D}_{\bar{1},\bar{9}}$ whose genitor is $7$. The set of all genitors in $\mathcal{D}$ will be denoted $\mathcal{C}_G(\mathcal{D})$.
To sum up, promoted, new and genitor CIs are defined as follows:\\ 
\begin{itemize*}
    \item $\mathcal{C}_P(\mathcal{D}) = \{ c ~|~ c \in \mathcal{C}(\mathcal{D}),~c \subseteq \iota(t_n) \}$,\\ 
%
%
    \item $\mathcal{C}_N(\mathcal{D}) = \{ \bar{c} ~|~ \exists c \in \mathcal{C}(\mathcal{D}),~\bar{c} = c \cap t_n, \kappa_{\mathcal{D}}(\bar{c}) \neq \bar{c} \}$,\\
%
%
    \item $\mathcal{C}_G(\mathcal{D}) = \{ c \in \mathcal{C}(\mathcal{D})~|~ c = \kappa_{\mathcal{D}}(c \cap t_n), c \not\subseteq t_n \}$. 

\end{itemize*}

Finally, the support in $\mathcal{D}^+$ for any $X$ in $\Delta(\mathcal{D}, t_n)$ is merely a unit more than the support of its closure in $\mathcal{D}$ (increase due to $t_n$). Indeed since 
$\sigma_{\mathcal{D}}(X) = \sigma_{\mathcal{D}}(\kappa_{\mathcal{D}}(X))$, 
we have 
$\sigma_{\mathcal{D}^+}(X) = \sigma_{\mathcal{D}}(\kappa_{\mathcal{D}}(X)) + 1$.
For instance, in Table~\ref{tab:ci_before_rem_second}, the support of the new CI $20$ is 4 while that of CI $7$, its genitor, is $3$.

Dually, let $t_o$ be an \textit{obsolete} transaction from $\mathcal{D}$ and $\mathcal{D}^- = \mathcal{D} - \{t_o\}$.
The impact of removing $t_o$ is some CIs, called \textit{obsolete}, vanishing in $\mathcal{D}^-$ while others, the \textit{demoted}, get their support decreased by 1. The corresponding sets of CIs are investigated in section~\ref{sec:design-theory}.

\section{Related work}
\label{sec:related}

Historically, methods for incrementally listing the closures of a cross-table date back at least to~\cite{lat:norris78}. In the 1990s and 2000s, a variety of intersection-based incremental concept lattice builders were published, starting with~\cite{galois_inc:godin+95} which introduced genitors. They compute jointly CIs with respective tidsets and precedence.
Later methods, e.g. \galiciat~and \galiciam~\cite{valtchev02}, reflect FCI mining concerns, thus they forgo precedence and tidsets
. However, they are bound to maintain all CIs as some genitors might well be infrequent. CIs are stored compactly, e.g. in prefix trees, and accessed trough inverted lists to avoid spending on empty intersections.

\cfistream~\cite{jiang2006cfi} is a sliding-window CI miner that heavily relies on intersections between CIs (as opposed to CI-to-$t_n$ ones). Its decrement lists all subsets of $t_n$ and finds their closures as intersections of all encompassing CIs to test in obsolescence; increments are more focused. Overall, genitors are targeted as such yet key properties thereof are ignored. 

\galiciap~\cite{Valtchev2008} is a landmark CI miner using inverted lists to selectively access CIs and trie storage. Intersection trie is grown item-wise with each CI pointing to its current prefix. Intersections are split into new/promoted by tracking the minimum generating CI. 

\clostream~is another intersection-based method, introduced as landmark and later on completed to a sliding-window mode~\cite{yen2011fast}. It uses inverted lists to filter CIs to get intersected with $t_n$/$t_o$. Genitors appear in both add and removal processing yet as two unrelated --and informally defined-- notions.

A landmark intersection-based approach is adapted to batch FCI mining in~\cite{borgelt2011finding}. They use a two-pass scheme and store nearly all CIs (stripped of infrequent items) and no tidsets. The approach reportedly outperformed modern batch FCI miners on specific types of data.

\moment~\cite{chi2004moment} is a sliding-window FCI stream miner adapting pattern enumeration (as in \textit{Eclat}~\cite{zaki97}) supported by a CE-tree.  Its increment proceeds by sibling node joins which exploit the non closed \textit{promising} itemsets plus some infrequent ones. Support is yielded by tidset intersections. Its decrement relies on direct closures computing to spot obsolete CIs.
\newmoment~\cite{li2006new} enhances \moment~with bit vector encoding of tidsets.
Yet it forgoes node categories: A node is joined with all its siblings.

\gctree~\cite{chen2007gc} adapts the batch FCI miner \dciclosed~\cite{lucchese2004dci} to mining CIs from streams. 
It uses a hybrid scheme: new CIs are generated either by  intersection of existing CIs or by the \textit{closure climbing}~\cite{lucchese2004dci}. Genitors are absent: Support is computed tidset-wise while the decrement exploits direct closure computing. Unlike \moment, at most one non closed itemset is kept for each class $[~]_{\mathcal{D}}$.

In summary, all methods incur overhead due to extra itemsets stored and inefficient new closure computing.
Moreover, while \clostream~and \galiciap~substantially limit the computation effort, they still suffer on sub-optimal memory usage, e.g. redundancy between CI storage and inverted lists. 

Our \ciclad~method aims at further minimizing both overheads.


\section{Overview of the approach}
\label{sec:design}

\ciclad~combines an incremental part that builds upon basic ideas from \galiciap~with a novel decremental method based on original mathematical results. Both share an overall computing schema, thus achieving high homogeneity and compactness. 
Below, we outline that schema and illustrate it (in incremental mode), then provide the formal background for decrementing and expand on high-level algorithmics.

\subsection{Approach summary}
\label{sec:design-over}

The generic online processing comprises three steps: (1) intersection computing; (2) splitting the total set into promoted (demoted) and new (obsolete) CIs; (3) update of the indexing structures on CIs. Below, we expand on each step while using generic notations, e.g. $t_x$ to mean $t_n$ or $t_o$. 

At step (1), intersection itemsets are grown from prefixes, along an iteration over all items in $t_x$. At an item $a_k$, each CI comprising $a_k$: (i) has its current prefix extended with $a_k$, and (ii) has its support checked against the current maximal support of a CI sharing that prefix.
Eventually, prefixes grow into complete intersections whereby each intersection $X$ has netted the minimum generating CI $\min_{\subseteq}([X]_{t_n})$.
Step (2) categorizes each $X$ by checking for a genitor CI within $[X]_{t_n}$: In \cicladA, the equality $X = \min_{\subseteq}([X]_{t_n})$ means $X$ is promoted, otherwise new. In \cicladD, the test is more subtle, as explained in section~\ref{sec:design-theory}.
Step (3) updates CI storage and inverted lists for items in $t_x$.

\subsection{Key techniques and data structures}
\label{sec:over-ex-inc}

On the algorithmic side, \ciclad~uses item-wise inverted lists for quick access to CIs and trie-based storage of evolving intersections (allows a unique copy per prefix). Thus, during step (1), a CI only keeps a pointer to the trie node with the last item of its current intersection prefix. At step (2), end nodes of full intersections in the trie are identified by the non-zero count of referring CIs. 

\begin{figure}[t]
\centering
     \includegraphics[width=0.85\linewidth]{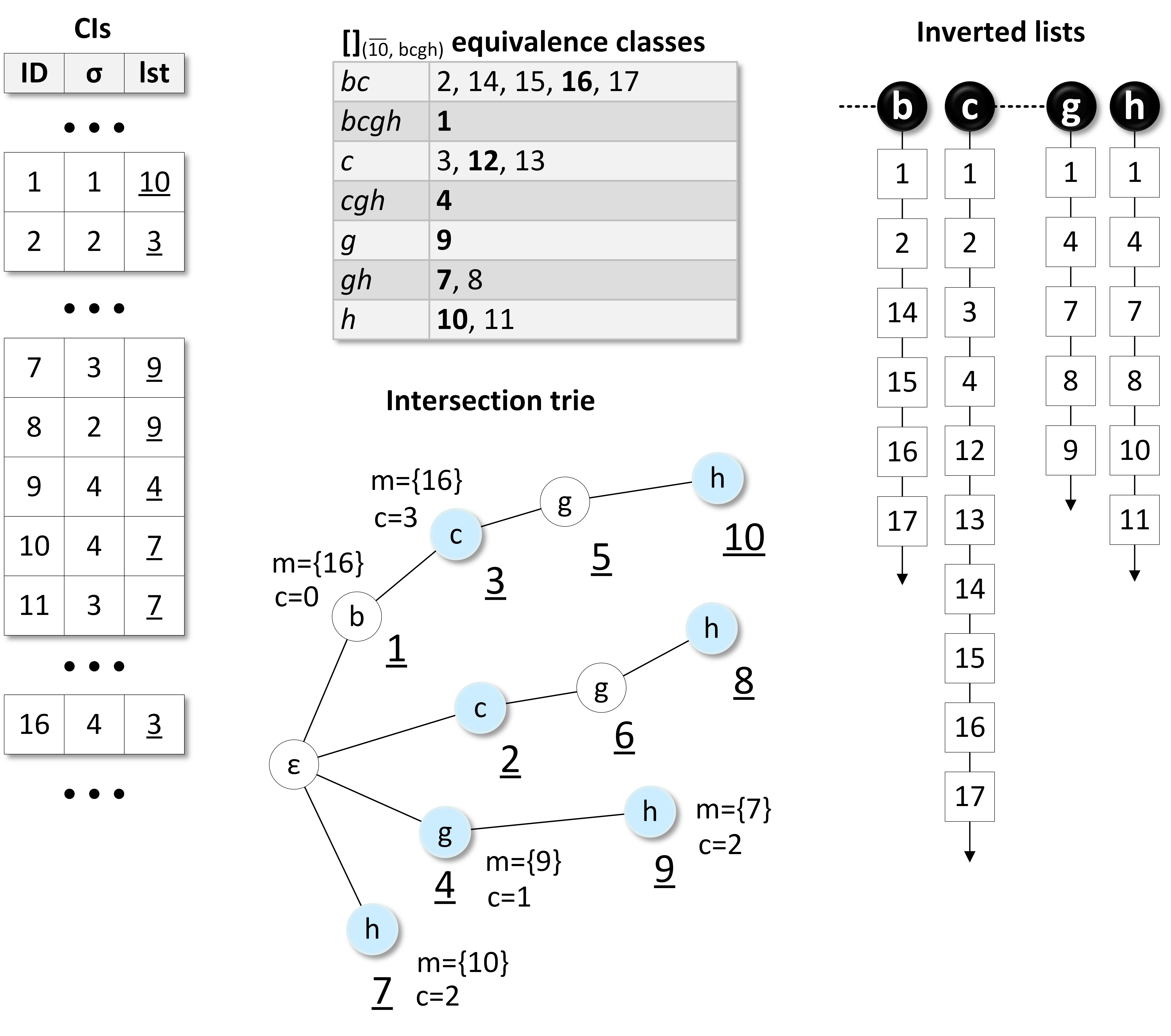}
      \caption{A snapshot of the working memory of \cicladA}
       \label{fig:highlevel}
\end{figure}

Figure~\ref{fig:highlevel} is a snapshot of the working memory of \cicladA~at the very end of the increment of $\mathcal{D}_{\bar{1},\bar{9}}$ (Table~\ref{tab:ci_before_tenth}) by $(\bar{10}, bcgh)$ (Table~\ref{tab:example_data}), i.e. after all four items have been processed. 
On the left, CI storage structure features fields for ID, support, and a reference to an end node (the $last$ field) in the intersection trie. 
Observe that itemsets are only stored in the inverted lists (on the right of Figure~\ref{fig:highlevel}), e.g. the ID of $7$ (valued $fgh$) appears in the lists of $f$ (not in the figure), $g$, and $h$.

The trie, which is built anew on each window shift, has nodes with fields for ID (underlined), the item, a pointer to the minimal CI (\textsl{min}) and a counter (\textsl{cpt}) for referring CIs.
The counters discriminate end nodes of full intersections (shaded in the figure) against the rest:
Due to a simple bookkeeping mechanism for shifting $last$ pointers, end nodes have at least one such pointer directed at them (hence a non-zero counter value). For instance, in $16$, whose intersection is $bc$, \textsl{last} points to $\underline{3}$ whereas $10$ and $11$ point to $\underline{7}$. Conversely, the \textsl{min} field of trie node $\underline{7}$ points back to $10$, the minimal CI yielding $h$. Overall, no white node is pointed at by a CI, hence they are ignored at step (2).

The trie grows 
along an iteration over items in $t_x$. For each item $a$, trie nodes with $a$ are appended to existing paths. To that end, CIs in the list of $a$ are scanned: For any such $c$, the node in $c.last$ is replaced with a successor carrying $a$ (if missing, such a node is created). The \textsl{cpt} fields of both former and new $last$ nodes are updated accordingly. Then, $c$ is tested for minimality w.r.t. the new $last$. Details of how \ciclad~updates the above fields are given in section~\ref{sec:design-code-common}. 

Finally, categorization tests equality of a node's intersection to its minimal CI
. In our example, equality holds for node $\underline{3}$ vs. CI $16$ ($16$ is thus promoted), but not for $\underline{7}$ vs. $10$, hence the intersection $h$ becomes a new CI ($19$ in Table~\ref{tab:ci_before_rem_second}). 
Overall, $\{bcgh\}$, $\{cgh\}$, $\{gh\}$ and $\{h\}$ are new CIs while $\{bc\}$, $\{c\}$, and $\{g\}$ are promoted. 

To sum up, the above structures jointly enable rapid computation of intersections and genitors while keeping a low memory footprint w.r.t. competitor methods.

\subsection{Decrement-related properties}
\label{sec:design-theory}

Dually to the incremental case, our decremental method computes $\mathcal{C}(\mathcal{D}^-)$ from $\mathcal{C}(\mathcal{D})$ and $t_o$, an \textit{obsolete} transaction in $\mathcal{D}$. Simply put, this amounts to yielding Table~\ref{tab:ci_before_tenth} from Table~\ref{tab:ci_before_rem_second} and $\bar{10}$ (though in a stream, $\bar{1}$ would vanish first).
By duality, $\mathcal{C}(\mathcal{D}^-) \subseteq \mathcal{C}(\mathcal{D})$, i.e. no new closures appear.
The relevant CI families are (1) \textbf{obsolete} CIs (to be removed from $\mathcal{C}(\mathcal{D})$), denoted $\mathcal{C}_O(\mathcal{D})$, and, (2) \textbf{demoted} CIs 'surviving' in $\mathcal{D}^-$ but with a decreased support, $\mathcal{C}_D(\mathcal{D})$. The situation is easily illustrated by mirroring our previous example. Thus, in Table~\ref{tab:ci_before_rem_second}, $20$ becomes an obsolete CI (genitor $7$) while $16$ is a demoted CI with support decreased from 5 to 4.

Notwithstanding apparent symmetry, an issue with decrements is $t_o$ does not discriminate obsolete vs. demoted CIs since both are subsets thereof:

\begin{theorem}
\label{prop:dec-int}
$\Delta(\mathcal{D}, t_o)$ correspond to obsolete and demoted CIs:
$\Delta(\mathcal{D}, t_o) = \{t_o \cap c ~|~ c \in \mathcal{C}(\mathcal{D})\} = \mathcal{C}_O(\mathcal{D}) \cup \mathcal{C}_D(\mathcal{D})$.
\end{theorem}

The above follows from Property~\ref{prop:fca-transactions-int} and $t_o \in \mathcal{C}(\mathcal{D})$. 
Now, while $\mathcal{C}_O(\mathcal{D}) \cup \mathcal{C}_D(\mathcal{D})$ stands out within $\mathcal{C}(\mathcal{D})$ via inclusion in $t_o$, further tests are needed to split it into components, e.g. presence of a genitor as a test for obsolescence.
Indeed, if $\mathcal{D}$ is seen as the 'increment' of $\mathcal{D}^-$ by $t_o$, then an obsolete $c_o$ becomes a new CI hence there should be a genitor $c_g$ in $\mathcal{D}$ s.t. $c_o = c_g \cap t_o$. Thus, $c_g$ the closure of $c_o$ in $\mathcal{D}^-$, i.e. $c_g = \kappa_{\mathcal{D}^-}(c_o)$. 
However, minimality of genitors, $c_g = \min([c_g]_{t_o})$, that proved crucial for incrementing, only holds in $\mathcal{D}^-$, but not in $\mathcal{D}$ where the minimum is $c_o$ ($c_o = c_o \cap t_o$ hence $c_o = \min([c_o]_{t_o})$). The genitor only becomes the minimum if non-inclusion in $t_o$ is also required, $c_g = \min([c_g]_{t_o} - \wp(t_o))$.
Albeit appealing (\clostream~went this way), extra non-inclusion tests may prove costly. 
Instead, we leverage the support difference of $c_o$ and $c_g$ in $\mathcal{D}$: For potential $c_o$ we look for $c_g \in [c_o]_{t_o}$ s.t. $\sigma_{\mathcal D}(c_o) =  \sigma_{\mathcal D}(c_g) + 1$. Such $c_g$ will nullify Definition~\ref{def:closed} for $c_o$ in $\mathcal{D}^-$. Formally:

\begin{theorem}
\label{prop:gen-dec}
A CI is obsolete \textit{iff} it has a genitor, $\mathcal{C}_O(\mathcal{D}) = \{c \in \mathcal{C}(\mathcal{D})~|~\exists c_g \in \mathcal{C}(\mathcal{D}) : c = c_g \cap t_o,~ \sigma_{\mathcal D}(c_g) =  \sigma_{\mathcal D}(c) - 1\}$.
\end{theorem}

The reasoning behind the \textit{if} part has been exposed in the previous paragraph. 
For the \textit{only if} part, given a CI $c \in \mathcal{C}(\mathcal{D})$, the above three-fold condition on $c_g$ must be shown to entail $c \not\in  \mathcal{C}(\mathcal{D}^-)$. First, $\sigma_{\mathcal D}(c_g) \neq  \sigma_{\mathcal D}(c)$ means $c_g \neq c$, hence $c_g \not\subseteq t_o$. 
Consequently, $\sigma_{\mathcal{D}^-}(c_g) = \sigma_{\mathcal D}(c_g)$, as $c_g$ is not impacted by $t_o$, while for $c$ it decreases by one, $\sigma_{\mathcal{D}^-}(c) = \sigma_{\mathcal D}(c) - 1$. Applying 3rd condition, $\sigma_{\mathcal{D}^-}(c_g) = \sigma_{\mathcal{D}^-}(c)$.
This and 2nd condition, $c \subseteq c_g$, imply $c \not\in \mathcal{C}(\mathcal{D}^-)$, hence its obsolescence.

To sum up, while a new CI is easy to spot among other intersections since smaller than its genitor CI, here an obsolete CI equals its intersection with ${t_o}$. Therefore categorization goes support-wise: The genitor of $c \in \mathcal{C}_O(\mathcal{D})$ is the CI in $[c]_{t_o}$, with only $t_o$ missing in its support set w.r.t. to $i_{\mathcal{D}}(c)$.

\subsection{Decrement-specific processing}
\label{sec:over-dec}

As explained above, for an intersection, its potential genitor, if any, has support one less than the minimal generating CI.
To make genitors emerge at step (1), a \textsl{gen} field is added to trie nodes to store candidate CIs.
Thus, whenever the intersection prefix of a CI $c$ is extended, $c$ is confronted to the current minimum of the node in its updated \textsl{last} field. If non minimal, $c$ is then compared to the content of \textsl{gen}. Observe that, due to the way intersections are grown, at some intermediate step more than one CI could satisfy the above criterion. Indeed, since inverted lists are not sorted, support values of CIs may come at arbitrary order. 
Next, minimal CIs being unique, there can be only one legit candidate in a \textsl{min} field at any point. However, the interplay between computation of \textsl{gen} and \textsl{min} fields requires a set of candidate CIs to be kept in both. 
Notwithstanding, ultimately a \textsl{gen} field can hold at most one CI satisfying Property~\ref{prop:gen-dec}.
The details of the resulting field updates are provided in section~\ref{sec:design-code-dec}.

The rest of step (1) mirrors the incremental case.
At step (2), discrimination should be as follows: The CI in \textsl{min} is demoted if \textsl{gen} field is empty, obsolete otherwise. However, we dropped removals from \textsl{gen} enforcing this condition to gain efficiency and instead check whether CIs are in $[c]_{t_o}$ (see Algorithm~\ref{alg:updateci_d} for details). Step (3) is the removal of obsolete CIs from the global CI storage, as well as from all inverted lists it appears in.

\section{CICLAD sliding-window miner}
\label{sec:design-code}

Common parts (superscripted by $^*$) of the overall computing schema are presented below followed by case-specific ones. Notice that to achieve homogeneity in step (1),
we initialize the above global structures with a special CI with id $0$, itemset $\mathcal I$ and support of $0$. 

\begin{algorithm}
  \caption{$Ciclad^{*}$}
  \label{alg:ciclad_generic}
  $trie \leftarrow init()$\\
  \ForEach{$a \in t_{x}.items$}{
  \label{lst:line:firsthalf}
	\ForEach{$c \in a.list$}{
	  \label{lst:line:listci}
		\If{$c.last = null$}{
		 $c.last \leftarrow trie.root$\\
         \label{lst:line:initroot}
         }
	  $ExpandPath^{*}(c, a)$\\
	  \label{lst:line:expandpath}
	  }
	  }
  \label{lst:line:updateci}
  $UpdateCIs^{*}()$\\
\end{algorithm}

\subsection{Unified schema}
\label{sec:design-code-common}

\ciclad$^{*}$~(Algorithm~\ref{alg:ciclad_generic}) is the high-level generic method to add/remove a transaction.
At step (1) it iterates over $t_x$ to yield its intersections with existing CIs~(lines~\ref{lst:line:firsthalf} to~\ref{lst:line:expandpath}).
For an item $a$, the current prefixes of all CIs (in $c.last$) of its inverse list ($a.list$) are extended by appending $a$~(line~\ref{lst:line:expandpath}). Prefixes are initialized to the root node (line~\ref{lst:line:initroot}).

\def\pp{\texttt{++}}
\def\mm{\texttt{{-}{-}}}
\def\tgen{t\texttt{-}gen}

\begin{algorithm}
\caption{$ExpandPath^{*}$}
  \label{alg:expand_path}
  
  
    $n \leftarrow lookup\_succ(c.last,a)$\\ 
  \If{$n = null$}
  {
  	$n \leftarrow create\_succ(a)$\\ 
  }
  $c.last.cpt\mm$; $n.cpt\pp$\\
  $c.last \leftarrow n$\\
  $UpdateGen^{*}(c)$\\
  \label{lst:line:updategen}
\end{algorithm}

$ExpandPath^*$~(Algorithm~\ref{alg:expand_path}) is the unit intersection step. From the $c.last$ node, it looks up the successor with $a$ (if none, creates it). Relevant fields are updated to reflect the extended prefix, before calling $UpdateGen^*$ for a case-specific bookkeeping of the top support CI(s).
Basically, each $a$ shared by a CI $c$ and $t_x$ pushes $c.last$ downwards in the trie by a node, up till the complete intersection is built.

\begin{figure*}[t]
\centering
     \includegraphics[width=\linewidth]{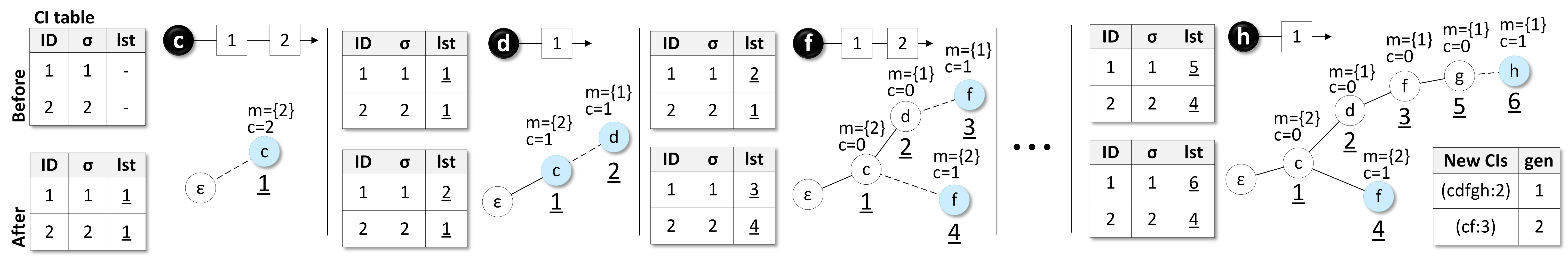}
      \caption{Trie evolution upon adding $\bar{3}$ to $\mathcal{D}_{\bar{1},\bar{2}}$ ($m$, $c$, and $lst$ stand for fields $min$, $cpt$, and $last,$ respectively).}
      \label{fig:add_trie_trx_4}
\end{figure*}

$UpdateCIs^*$ covers the above steps (2) and (3).
Within the final trie, it filters complete intersections and categorizes them. The CI storage update is case-specific (see $UpdateCIs^+$ and $UpdateCIs^-$).
\begin{algorithm}
\caption{$UpdateCIs^{+}$}
  \label{alg:updateci_p}
\ForEach{$n \in nodes(trie)$}
{
	\If{$n.cpt \neq 0$}
	{
		\label{lst:line:secondhalf}
    	\eIf{$n.depth = n.min.size()$}
		{
	  	$n.min.supp\pp$\\
	  	\label{lst:line:modified}
		}
		{
     	\label{lst:line:new}
	  	$createCI(path(n), n.min.supp + 1)$ \\
		}
  	}
}
\end{algorithm}

\subsection{Increment-specific processing in \ciclad$^{+}$}
\label{sec:design-code-inc}

Update minimal CIs ($UpdateGen^{*}$) is the first step to differentiate both cases. $UpdateGen^{+}$ (Algorithm~\ref{alg:expand_path},  line~\ref{lst:line:updategen}) merely compares current CI $c$ and CI in $n.min$ support-wise and updates the latter. We skip it here since straightforward.

$UpdateCIs^+$~(Algorithm~\ref{alg:updateci_p}) looks up the final trie for intersection end nodes and discriminates them with a cardinality test. A new intersection is recognized by its size (end node's depth) being smaller than the size of the CI in $n.min$ (stored for all CIs).
Then, $createCI()$ implements step (3): It pushes the new CI into the inverted lists of all its items (retrieved from the trie path by $path(n)$). In case of promoted CI, step (3) is a mere support increase.

Figure~\ref{fig:add_trie_trx_4} traces the evolution of the trie upon inserting $\bar{3}$ 
into $\mathcal{D}_{\bar{1},\bar{2}}$ 
whereby $\mathcal{C}(\mathcal{D}_{\bar{1},\bar{2}}) = \{(\bar{1}, abcdefgh), (\bar{2},abcef) \}$.
Each section of the figure shows the impact of a single item: The inverted list is on top, the changes in the $last$ column of the CI table on the left, and the current trie on the right. In the trie, nodes are decorated by current minimal CI and count of all referring CIs.

\begin{algorithm}
\caption{$UpdateGen^{-}$}
  \label{alg:update_gen_d}
$\textbf{switch}~(c.support - c.last.max\_supp)$\\
 ~~$\textbf{case}~\geq 2$ : $c.last.min \leftarrow \{c\}$;~$c.last.gen \leftarrow \varnothing$\\
 ~~$\textbf{case} ~~~1$ : $c.last.gen \leftarrow c.last.min$;~$c.last.min \leftarrow \{c\}$\\
 ~~$\textbf{case} ~~~0$ : $c.last.min \leftarrow c.last.min \cup \{c\}$\\
 ~~$\textbf{case}~-1$ : $c.last.gen \leftarrow c.last.gen \cup \{c\}$\\
$c.last.max\_supp \leftarrow \max(c.supp, c.last.max\_supp)$\\
\end{algorithm}

\subsection{Decrement-specific processing in \ciclad$^{-}$}
\label{sec:design-code-dec}

$UpdateGen^{-}$ (Algorithm~\ref{alg:update_gen_d} below) covers the joint maintenance of the $min$ and $gen$ fields of trie node, i.e. the minimal CI and the 'less-by-one' candidates, respectively. As indicated above, both store sets of CIs.
Given a CI $c$ whose current intersection prefix end has been freshly redirected to a trie node $n$, the following reasoning applies to its own support of that CI and the maximal support stored at $n$ ($max\_supp$ field): If the support of $c$ is way higher (2+), then both fields are flushed and $c$ becomes the new minimum, whereas with a difference of one, the current $min$ is merely shifted to $gen$. With 0 or -1, $c$ is added to $min$ or $gen$, respectively. Other values trigger no action. At the end, the maximal support for $n$ is updated correspondingly.

\begin{algorithm}
\caption{$UpdateCIs^{-}$}
  \label{alg:updateci_d}
\ForEach{$n \in nodes(trie)$}
{
	\If{$n.cpt \neq 0$}
	{
		$demoted \leftarrow true$\\
		\ForEach{$c \in n.gen$}
		{
			\If{$c.last = n$}
			{
				$removeCI(n, n.min)$\\				
				$demoted \leftarrow false$\\
				$break$\\
			}
		}
		\If{$demoted$}{
			$n.min.supp\mm$\\
		}
  	}
}  
\end{algorithm}

\begin{figure*}[t]
\centering
     \includegraphics[width=0.85\linewidth]{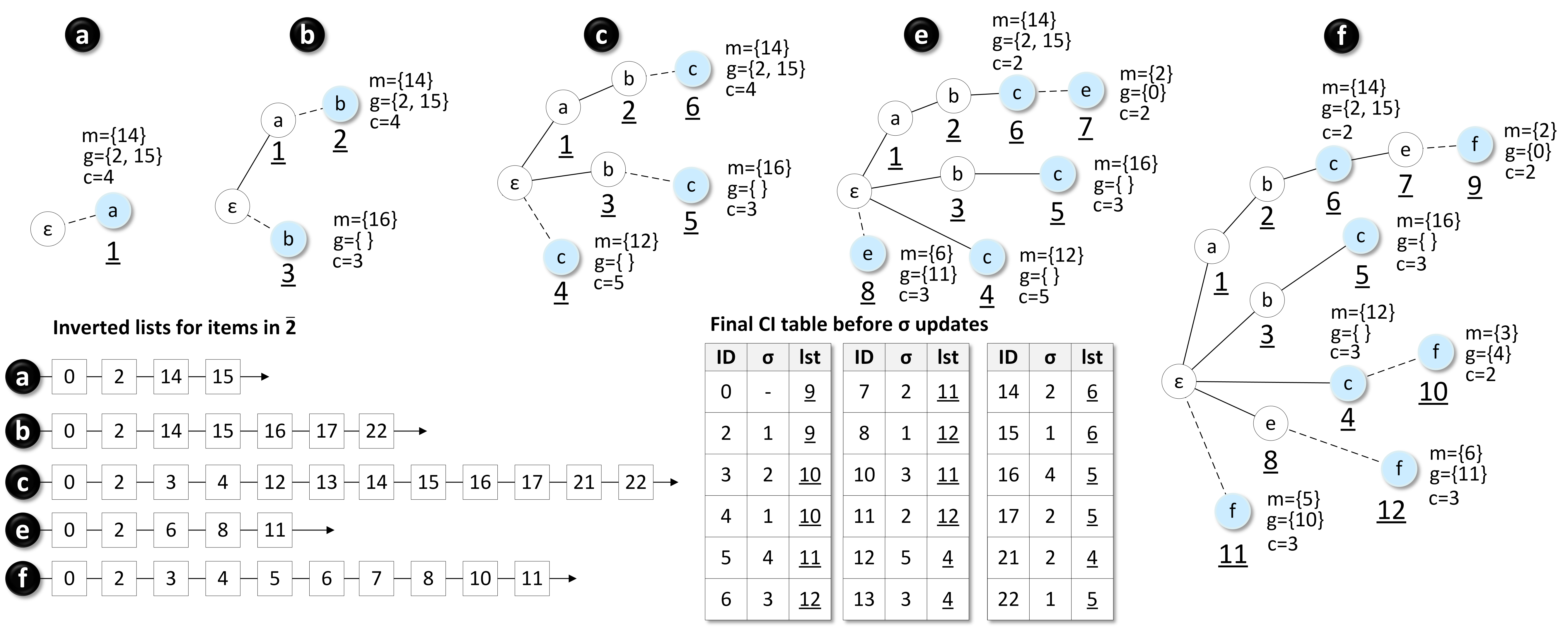}
      \caption{Trie evolution upon removing $\bar{2}$  from $\mathcal{D}_{\bar{2},\bar{10}}$ ($m$, $g$, $c$ and $lst$ stand for fields $min$, $gen$, $cpt$ and $last$, respectively).}
       \label{fig:rem_trie_trx_2}
\end{figure*}

Within the final trie, $UpdateCIs^{-}$ covers step (2) and (3), hence it first categorizes full intersections and updates the CI family based on respective $min$ and $gen$ fields of their end nodes. In actuality, while proper maintenance of $gen$ would ensure that eventually at most one CI is stored at each node, we decided to drop $gen$ updates upon moving the $last$ pointer of a CI away from the node. 
Thus, a CI may belong to $gen$ lists of end nodes corresponding to strict prefixes of its intersection. This is illustrated by part $e$ of Figure~\ref{fig:rem_trie_trx_2}: the CI 2 ($abcef$) remains in the $gen$ set of the end node of $abc$, although its $last$ field eventually points to the end of $abcef$ (see part $f$ of Figure~\ref{fig:rem_trie_trx_2}). Property~\ref{prop:dec-int} (see section~\ref{sec:design-code-rem}) guarantees that only CIs from $\Delta(\mathcal{D}, t_o)$ can be left behind in prefixes' $gen$ fields in the way described above. To detect them within $gen$, we check for $last$ fields pointing to a different trie node.

Conversely, genitor test spots CIs in $n.gen$  whose $last$ points to $n$. 
If positive, the test triggers $removeCI$ (step (3)) that removes the CI in $n.min$ from the inverted lists of all its items.

As an illustration, consider the removal of transactions $\bar{1}$ and $\bar{2}$ from the CI family in Table~\ref{tab:ci_before_rem_second}.
The first one is trivial as the only obsolete is $1$: All CIs being subsets of $\bar{1}$, only the special CI $0$ can be the \textit{genitor}. Thus, the resulting CI family is readily derived from Table~\ref{tab:ci_before_rem_second} by decreasing supports by 1.
Item iterations in the removal of $\bar{2}$, or $abcef$, (see Algorithm~\ref{alg:ciclad_generic}) are shown in Figure~\ref{fig:rem_trie_trx_2}. In the final trie (item $f$), intersections $\{abc\}$, $\{cf\}$ and $\{ef\}$ correspond to obsolete CIs. Indeed, each $gen$ field in the respective end node refers to a CI whose $last$ field, in turn, points to that node. 
For instance, the $gen$ field for $\underline{10}$ ($\{cf\}$) contains $4$, or $\{cdfgh\}$, of support 1 while the $min$ field value is $3$ $\{cf\}$ of support 2). The demoted CIs are $\{bc\}$ and $\{c\}$ which have both an empty $gen$ field.



\section{Evaluation}
\label{sec:eval}

We compared experimentally \ciclad~to \moment, \newmoment, \clostream~and \cfistream. 
The original version of \textit{Moment} was used as provided by its authors. To level the playing field, we implemented \ciclad~in C++ and in a single threaded mode\footnote{Code available at \url{https://github.com/guyfrancoeur/ciclad}} as well as \newmoment, \clostream, \cfistream\footnote{While the original versions  would have been preferable, \textbf{none} of these is currently made \textbf{available} by respective authors.}. To investigate their relative efficiency, 
we put the methods in identical conditions, i.e.  we made them compute \textit{all CIs} from varying datasets and sliding window lengths. 

\subsection{Datasets}
\label{sec:eval-data}

We used seven datasets of varying nature (see metrics in Table~\ref{tab:datasets}). %
\textit{Mushroom} and \textit{Retail}, are popular datasets\footnote{\url{http://fimi.ua.ac.be/data/}}:
\textit{Retail} is a sparse market basket dataset, while
\textit{Mushroom}, describing mushroom samples, is a dense and correlated dataset made of same-size transactions.
\textit{Synth} and \textit{Synth2} are synthetic transactional datasets, generated with SPMF\footnote{\url{http://www.philippe-fournier-viger.com/spmf/}}, of medium and small size, respectively.
Three other real-world datasets were used: click streams for \textit{BMS-View} (from \text{KDD 2000}), online purchases for \textit{Chainstore} (from the \textit{Nu-MineBench} project), and network logs\footnote{\url{https://gitlab.com/adaptdata/e2}} adapted from activity data from the \textit{DARPA Transparent Computing} program\footnote{\url{https://www.darpa.mil/program/transparent-computing}}.

\begin{center}
\small
\begin{tabular}{|l|c|c|c|c|c|c|}
\hline
	\textbf{Dataset} & \textbf{$|\mathcal{D}|$} & $avg(|t|)$ & \textbf{$|\mathcal{I}|$} & \textbf{$stdev{(|t|)}$} & Density\\
\hline
\textit{Retail} & 88163 & 10.4 & 16470 & 55.9 & 0.06\%\\
\hline
\textit{Mushroom} & 8124 & 23 & 119 & 0 & 19.3\%\\
\hline
\textit{Synth} & 100000 & 25.4 & 10000 & 14.4 & 0.25\%\\
\hline
\textit{Synth2} & 10000 & 25.5 & 1000 & 14.3 & 2.5\%\\
\hline
\textit{BMS-View} & 77512 & 4.6 & 3340 & 6.1 & 0.13\%\\
\hline
\textit{Chainstore} & 1112949 & 7.2 & 46086 & 8.9 & 0.015\%\\
\hline
\textit{Net-Log} & 272376 & 6.1 & 299 & 3.7 & 2.04\%\\
\hline
\end{tabular}
\captionof{table}{Description of the datasets}
\label{tab:datasets}
\end{center}

\subsection{Experimental settings}
\label{sec:eval-settings}

All experiments were ran on Windows 10 Professional 64 bits  with Intel i7-8700 CPU and 32 GB of RAM.
We measured the total CPU time over the entire stream and set a time limit of 10K secs: Methods that ran longer on a dataset were aborted, while recording memory usage, and withdrawn from experiments on larger windows over the same dataset.
Moreover, we recorded peak memory usage rather than average across all windows.

Noteworthily, finer measures, like the evolution of time/memory values as well the number of CIs along the stream, albeit potentially helpful, could not be provided here for space reasons (e.g. see section~\ref{sec:results-detailed-disc}).

\begin{figure*}[ht]
\centering
     \includegraphics[width=\linewidth]{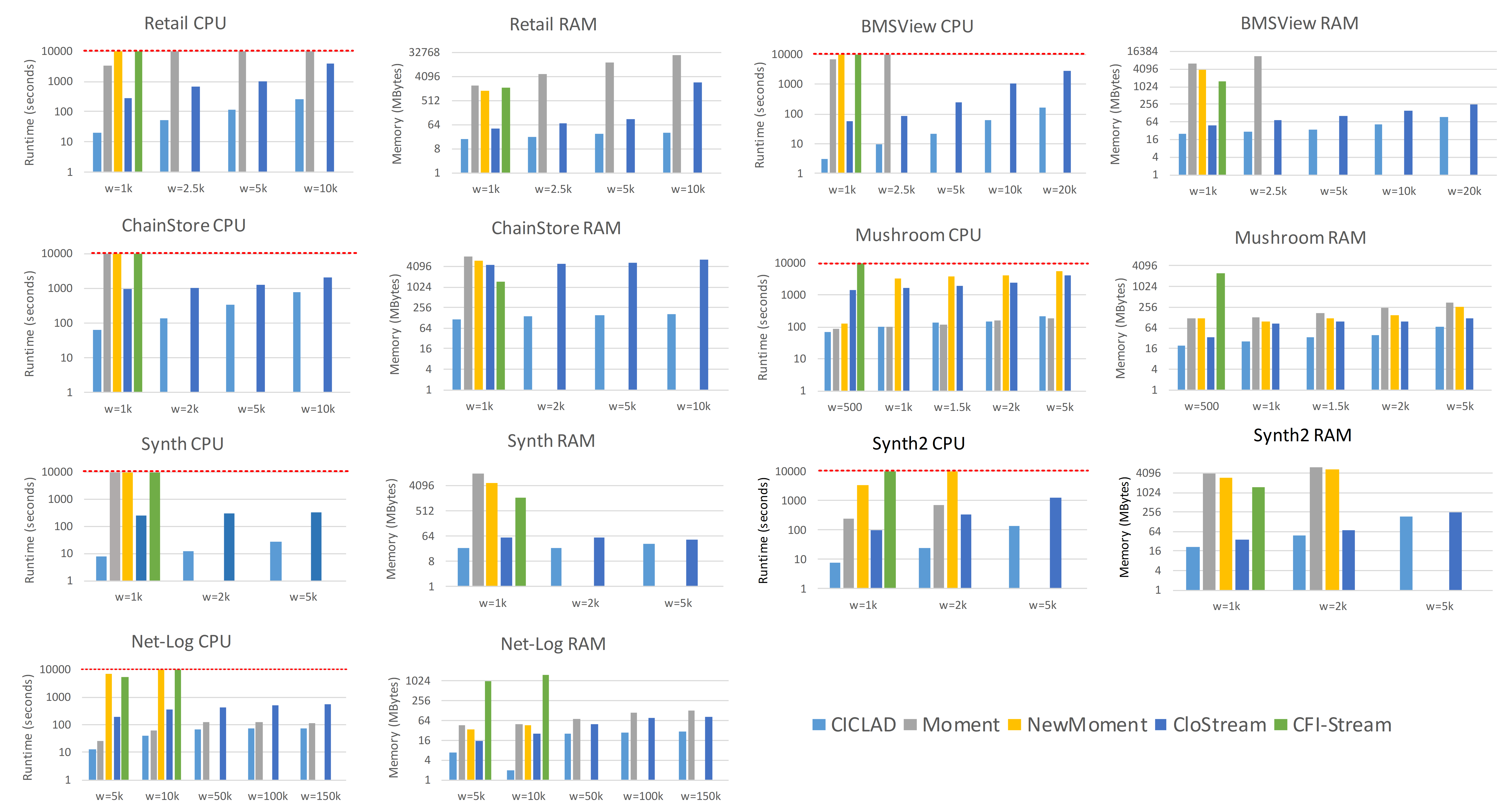}
      \caption{CPU time and memory usage of \ciclad, \moment, \newmoment, \clostream, and \cfistream}
       \label{fig:all_results}
\end{figure*}

\subsection{Results}
\label{sec:eval-result}

Results are summarized in Figure~\ref{fig:all_results} whereby ${w}=W$ indicates window size while the dashed line on the top is the time limit.

As a general trend, \cfistream~exceeded the time limit on the smallest window of every dataset, except \textit{Net-Log}, followed closely by \newmoment~which could process all window sizes only on \textit{Mushroom}. 
Next came \moment: It went off-limit on the first window size of \textit{Synth} and \textit{Chainstore}, and on the second one for \textit{BMS-View}, \textit{Synth2} and \textit{Retail} (yet, exceptionally, we let it run on all others and measured memory usage). On \textit{Mushroom}, it performed slightly better time-wise, yet its memory usage was substantially worse (among successful methods). 
Finally, \clostream~and \ciclad~finished within limit overall, whereby \ciclad~showed invariably better time and memory figures.

On the dataset side, it is noteworthy how on \textit{Retail}, with increasing window sizes, \moment~gradually approaches the limit of 32 GB of RAM, while \ciclad~remains very competitive (39 MB) and \clostream~less so.
The data in \textit{BMS-View} is similar to \textit{Retail} with less items and smaller transactions which enabled larger windows, e.g. of size 20K. It showed similar overall pattern, yet with a smaller gap between \ciclad~and \clostream~on RAM. On \textit{Chainstore} the same trend was observed, yet with larger gaps on memory. This is only half a surprise as both represent real-world streams. The  
\textit{Synth} and \textit{Synth2}, complete the picture of the dominance of \ciclad~and, to a lesser degree, \clostream. 

With \textit{Mushroom}, a very dense dataset, the trend is different: CPU-wise, there seems to be a tie between \ciclad~and \moment, far ahead of \clostream~followed by \newmoment.
On memory usage, it is less clear, yet \ciclad~is somewhat ahead of the rest. This is matched almost perfectly by the pattern on \textit{Net-Log}, despite the size of the windows being much larger. The only difference here is \newmoment~not being among the contendors after its timeout in the second window.

\subsection{Discussion}
\label{sec:eval-discuss}

From the above observations, we conclude that on storage intersection-based methods (\cfistream~excluded) perform invariably better than pattern enumeration ones. We believe this is the impact of storing non closed itemsets in each $[~]_{\mathcal D}$ class. This impact deepens with sparse data as classes tend to grow larger due to the ratio FCI/FI.

On sparse data, intersection-based methods are also faster than their competitors. Again, it is the overhead of non closed itemsets: Upon each CI lookup, \moment~traverses its $[~]_{\mathcal D}$ class from a smallest member up to the largest one (the CI) by walking along a chain of \textit{intermediate} itemsets. As a result, on sparse datasets (e.g. $\textit{Synth2}$) with limited-size windows \ciclad~can be up to 40 times faster than \moment. 

With dense datasets, pattern enumeration methods are favored as the FCI/FI ratio is higher, hence the smaller $[~]_{\mathcal D}$ classes. Conversely, the intersection-based equivalence classes $[~]_{t_{x}}$ tend to grow larger which increases the intersection effort per CI in intersection-based methods.
However, it is also worth recalling that with such data, the benefit of mining FCIs, as opposed to plain FI, is limited. 

Finally, \clostream~lags behind \ciclad~because of its fully-blown intersection operations and recurrent lookups for an intersection $X$ each time it is generated. \ciclad~streamlines both operations by its item-wise trie-based intersection growing technique.
\section{Conclusion}
\label{sec:conclude}

Our novel sliding-window miner \ciclad~implements an efficient intersection-based computing schema. It exploits the mathematically-grounded notion of genitor, the CI that is the closure of a non closed itemset which changes its status w.r.t. closedness upon window shift.
Design pillars in our intersection-based scheme include per-item inverted list storage of CIs, item-wise intersection growth, and support-based genitor detection.
The outcome of our experimental study confirmed that 
\ciclad~outperforms its competitors significantly, both on storage and processing.

As a basis for further research, \ciclad~lays the groundwork for additional challenges to be taken up. For instance, to tackle the mining of strong AR, or rather condensed representations thereof, over a sliding window, we are designing extensions thereof covering generator itemsets and/or precedence links as proposed in~\cite{nehme05oncomputing}. As a separate track, we investigate mining of rare yet confident AR~\cite{szathmary2010generating} from the stream. Next, we plan to leverage \ciclad's homogeneity in merging of $t_n$ and $t_o$ processing in a single-pass method.
Finally, as a way to focus strictly on FCIs, we will look at the evolution of the FI border~\cite{gunopulos_1997_data,karim2018mining}. 

\section*{Acknowledgments}
Thanks go to Y. Chi for the code of MomentFP and S. Benabderrahmane for the pre-formatted DARPA data.

\bibliographystyle{plain}
\bibliography{biblio}


\newpage


,
\section*{Appendix}

Below, we provide additional results about \ciclad~that clarify aspects such as its correctness, computational cost, and the way it compares to \gctree, a method that was excluded from the final validation study.

\section{Correctness results}
\label{sec:design-code-rem}

To show \ciclad~is correct, we first examine the decision about where intersections end.
Let $T$ denote the final trie and $n$ a node in $T$. Now, $n$ is the end of a full intersection, i.e. the itemset $items(n)$ made of items along the root-bound path from $n$ is in $\Delta(\mathcal{D}, t_x)$, iff its counter $n.cpt$ is strictly positive:
\begin{theorem}
\label{prop:trie-end}
Given a node $n$ in the trie~$T$, $items(n) \in \Delta(\mathcal{D}, t_x)$ iff $n.cpt > 0$.
\end{theorem}

Consider the evolution of the minimal CIs for a trie node.
Understandably, we only focus on trie states at the end of a particular iteration, i.e. with the respective item inverted list fully parsed. Assume after the $k$-th iteration, the list of item $a_k$ is processed and yielded a (still partially completed) trie $T_k$. Let $T_k[a_k]$ denote the set of nodes labelled by $a_k$. These nodes represent the increment w.r.t. $T_{k-1}$.

\begin{theorem}
\label{prop:trie-min}
Given a node $n \in T_k[a_k]$, and a CI $c \in {\mathcal C}({\mathcal D})$, $n.min = c$ iff $c = \kappa_{\mathcal D}(items(n))$.
\end{theorem}

Noteworthily, Property~\ref{prop:trie-min} ensures that in the final trie, all the minimal CIs are correctly positioned. For the decrement case, we need to further show that within any $gen$ field, at most one CI is not a subset of $t_o$. 

\begin{theorem}
\label{prop:trie-gen}
Let $n$ be a node in the final trie $T_{|t_o|}$ of an obsolete transaction $t_o$, 
then $|n.gen - \Delta(\mathcal{D}, t_o)| \leq 1$.
\end{theorem}

This follows from Property~\ref{prop:gen-dec} and the observation that CIs outside $\Delta(\mathcal{D}, t_o)$ keep their supports from ${\mathcal C}({\mathcal D})$ in ${\mathcal C}({\mathcal D}^-)$. Indeed, assuming more than a single CI satisfies the conditions, entails there are two different CIs in ${\mathcal C}({\mathcal D}^-)$ with support equal to the support of the obsolete/demoted itemset in $n.min$. This, regardless of the exact status of that itemset in $\Delta(\mathcal{D}, t_o)$, is a contradiction. To sum up, the $n.gen$ field of a node in the final trie can contain at most one CI outside $\Delta(\mathcal{D}, t_o)$ plus a number of CIs from that set. Then, only the former belongs to the class of $n$ (with minimum CI $n.min$).

Finally, the status of a CI $c$ in $n.gen$ depends on its being member of $\Delta(\mathcal{D}, t_o)$. To avoid costly tests of inclusion into $t_o$ we rely on the intersection class of $c$: since all CIs from  $\Delta(\mathcal{D}, t_o)$ are minimal in their own classes, each has a unique value in its $last$ field. Thus, none of the CIs $c$ in a $n.gen$ field that is also in $\Delta(\mathcal{D}, t_o)$, could refer to $n$ via its $c.last$ field:

\begin{theorem}
\label{prop:trie-test-obs}
Let $n$ be a node in the final trie $T_{|t_o|}$,
then for $c_g \in n.gen$, $c_g \not\in \Delta(\mathcal{D}, t_o)$ iff $c_g.last = n$.
\end{theorem}

\section{Complexity analysis}

The window shift complexity of \ciclad~is $O(k_m * l_m)$ in time and $O(k^{2}_{m} * l_m)$ in space.
Here, $k_m$ is the maximal transaction (and CI) size and $l_m$ the maximal number of CIs in a window.

The intersection computing (\ciclad$^{*}$ up till the end of \textit{ExpandPath}$^{*}$) is in $O(k_m * l_m)$.  
For each item $i$ from $t_x$ and CI $c$ comprising $i$, \ciclad$^{*}$ pushes the intersection of $c$ down its path in the trie. This involves few operations (half a dozen) all of constant time cost.

Categorizing intersections and creating new CIs (\textit{UpdateCIs} up till \textit{createCI()}) has also a cost in $O(k_m * l_m)$. 
First, detecting end nodes is in $O(k_m * l_m)$ since there are at most $l_m$ intersections, each of size at most $k_m$, hence $O(k_m * l_m)$ nodes in the trie. Next, creating all new CIs is also in $O(k_m * l_m)$: the same number bound $l_m$ multiplied by the unit cost of creation (linear in $k_m$).
Inverted lists can be updated in $O(k_m * l_m)$ time as each combination of a new CI and incident item amounts to one list insertion. 

\ciclad$^{*}$ has a memory footprint in $O(k^{2}_{m} * l_m)$. Indeed, the intersection trie will have at most $k_m * l_m$ nodes (see above) whose successor structures might need up to $k_m$ memory cells each.
Comparatively, the total footprint of the inverted lists is in $O(k_m * l_m)$.
Now, $k_m * l_m$ is, in fact, a gross overestimation of \textit{the total number of items in all CIs} which is key cost factor in both time and memory: The real figure, especially with sparse data, will be way lower. Noteworthily, the size of the window is not a factor in the above functions: This is the effect of skipping tidsets altogether (yet it has an indirect impact through $l_m$).

 Finally, \ciclad~is a listing algorithm, hence to be assessed not by total time but rather by the cost per output element, i.e. CI. Thus, assuming the entire stream processing cost is in $O(n_s * k_m * l_m)$, where $n_s$ is total number of transactions, the per-CI cost is, grossly, in $O(n_s * k_m)$, i.e. a polynomial in the size of the dataset. In contrast, the cost of producing a particular new CI $c$ in \moment~might go beyond that limit as the number of unpromising nodes traversed while generating $c$ can grow up to exponential in its size.

\section{Additional performance tests}
\label{sec:results-detailed-disc}

\subsection{Fine-grained performance analysis}
\label{sec:results-detailed-ciclad-moment}

\begin{figure}[ht]
\centering
     \includegraphics[width=.9\linewidth]{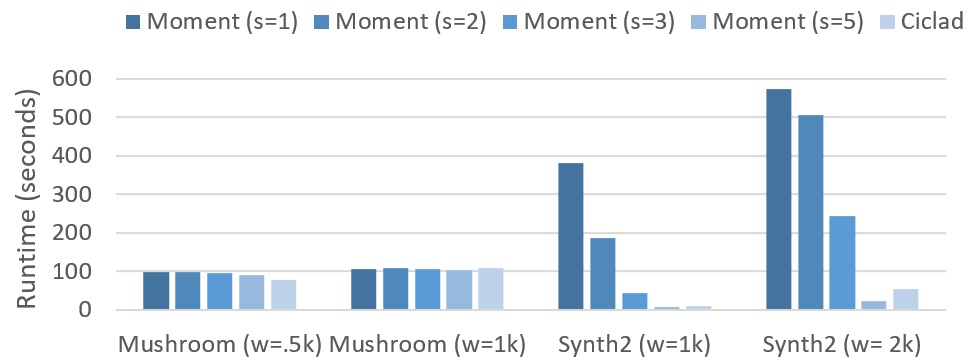}
      \caption{Runtime comparison varying $Moment$'s $min\_supp$}
       \label{fig:ciclad_vs_moment_minsupp_runtime}
\end{figure}

\begin{figure}[ht]
\centering
     \includegraphics[width=.9\linewidth]{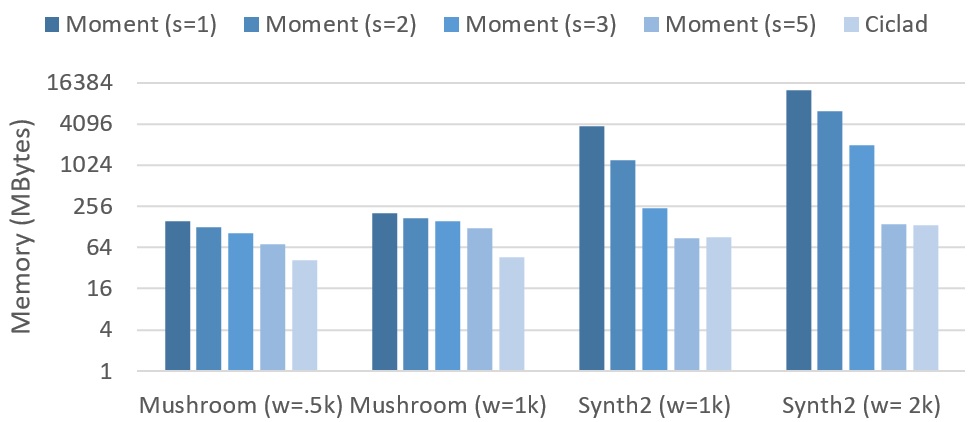}
      \caption{Memory comparison varying $Moment$'s $minsupp$}
       \label{fig:ciclad_vs_moment_minsupp_memory}
\end{figure}

We made \ciclad~compete on \moment's terms, i.e. with higher support thresholds. Thus, we compared both over various \textit{min\_supp} values (1,2,3 and 5), this time using only two datasets, one dense (\textit{Mushroom}) and one sparse (\textit{Synth2}), each with two different window sizes.
The results are summarized in Figure~\ref{fig:ciclad_vs_moment_minsupp_runtime} (CPU time) and in Figure~\ref{fig:ciclad_vs_moment_minsupp_memory} (memory usage).

On \textit{Mushroom}, variations in \textit{min\_supp} modestly impact the computing time of~\moment; this arguably fits the intuition that CIs are more regularly scattered over the pattern space (thus higher values are needed for a palpable drop in the cost).
Noteworthily, \ciclad~and \moment~offer comparable performances. On \textit{Synth2}, \moment's runtime efficiency improves much faster and it outperforms \ciclad~for thresholds of 5 and above.

Memory-wise, \ciclad~is still somewhat ahead, yet the trend of rapidly decreasing consumption in  \moment~is visible. Again, for the sparse dataset, with thresholds of 5 and up, \moment~catches up with \ciclad, whereas with the dense data the break-even point is still somewhere above.

\begin{figure}[ht]
\centering
     \includegraphics[width=\linewidth]{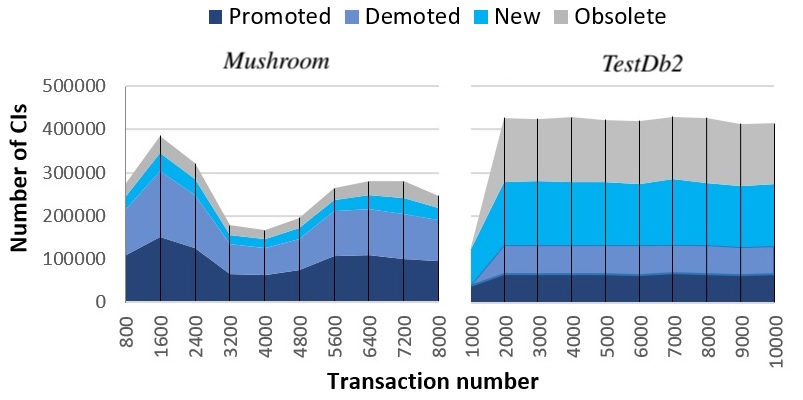}
      \caption{Evolution of $\mathcal{C}(\mathcal{D})$ in $Mushr._{w=1k}$ and $Synth2_{w=1k}$}
       \label{fig:modified_and_new_mushroom}
\end{figure}

As a possible hint at the reasons behind the observed performances, we track the proportion of \textit{new}, \textit{promoted}, \textit{demoted} and \textit{obsolete} CIs in windows. Results for \textit{Mushroom} and \textit{Synth2} datasets with windows of size 1K are shown in Figure~\ref{fig:modified_and_new_mushroom}. In summary, the higher ratio of new/obsolete CIs to promoted/demoted ones in sparse data would explain superior performances of \ciclad~by the costly tree restructuring in \moment~as opposed to inexpensive updates of existing nodes. Conversely, it hints at detecting of promoted/demoted CIs in \ciclad~as possible improvement point for speeding up dense data processing.

\begin{center}
\footnotesize
\begin{tabular}{|l|c|c|c|c|}
\hline
	\textbf{Dataset} & \textbf{Avg. nodes} & \textbf{Avg. CIs} & \textbf{Ratio}\\
\hline
\textit{Mushr. (w=1k,s=1)} & 677202 & 208952 & 3.24\\
\hline
\textit{Mushr. (w=1k,s=5)} & 234949 & 30081 & 7.08\\
\hline
\textit{Mushr. (w=.5k,s=1)} & 459379 & 16428 & 28.67\\
\hline
\textit{Mushr. (w=1.5k,s=1)} & 855548 & 44803 & 19.50 \\
\hline
\textit{Synth2 (w=.3k,s=1)} & 1655106 & 8913 & 184.78 \\
\hline
\textit{Synth2 (w=1k,s=1)} & 16013671 & 80364 & 199.26 \\
\hline
\textit{Synth2 (w=2k,s=1)} & 62125880 & 267717 & 232.21 \\
\hline
\end{tabular}
\captionof{table}{Average number of CIs and CET nodes in \moment}
\label{tab:ratio_ci_cetnodes}
\end{center}

We also examined the storage overhead in \moment, i.e. due to the storage of promising and intermediate itemsets. Table \ref{tab:ratio_ci_cetnodes} shows the average number of nodes within \moment's CE-tree (\textbf{Avg. nodes}) against the average number of CIs (\textbf{Avg. CIs}), both taken over the entire stream, for a number of combinations (dataset, window size, min\_supp). 
The wide variation, $3.24$ to an extreme $232.21$, is intriguing. Yet the trend correlates with the observations on computing time and memory usage, i.e., the higher the value, the less competitive the method vs \ciclad.

\subsection{\gctree~vs \ciclad~vs \moment}
\label{sec:results-detailed-ciclad-moment}
We studied also \gctree~in order to assess its hybrid approach. However, its decremental part was impossible to implement due to inconsistencies in the description of the method. 
Therefore, \gctree~was compared to \ciclad~and \moment, in \textbf{landmark} mode only.  

\begin{figure}[ht]
\centering
     \includegraphics[width=\linewidth]{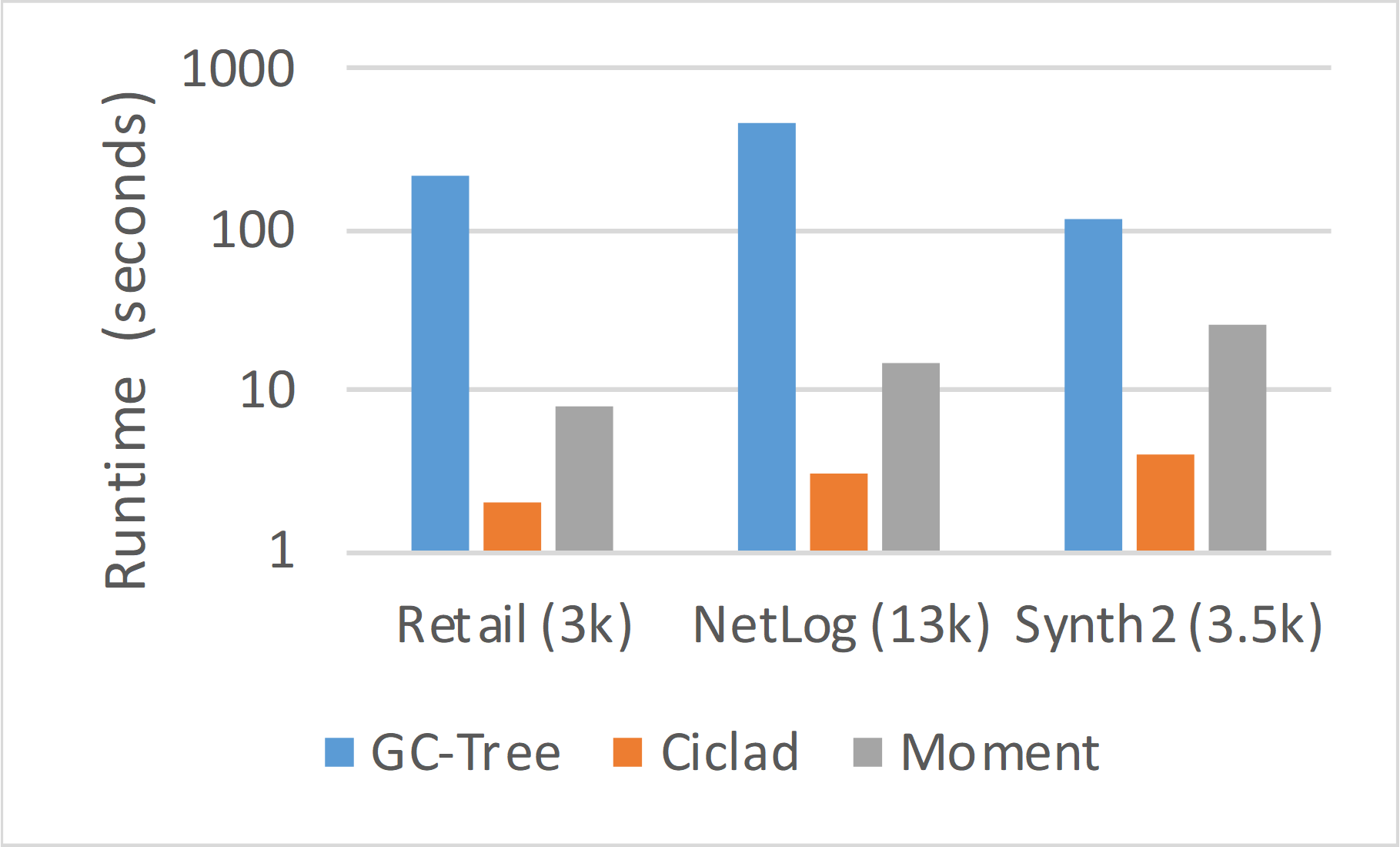}
      \caption{\gctree~vs. \ciclad~vs. \moment~in landmark mode}
      \label{fig:gctree_vs_others}
\end{figure}

In Figure~\ref{fig:gctree_vs_others}, an extract of the performance tests is given: The figure presents the CPU time on three of the seven datasets in landmark mode. We used a prefix large enough to let a stable trend appear.

An immediate observation is that the hybrid approach, even if appealing, does not perform well with large number of items. 
The clear gap between \gctree~and its competitors is, we surmise, due to the number of canonicity tests it needs to perform while extending a closure in order to ensure that the result is indeed the lexicographically smallest among all alternative extensions in its equivalence class $[~]_{\mathcal D}$.

\end{document}